\RequirePackage{etoolbox}
\csdef{input@path}%
{%
 {sty/},%
}
\documentclass[preprint,10pt,3p, twocolumn, number]{elsarticle}

\usepackage{amssymb}

\usepackage{amsmath}

\usepackage[hyphens]{url}
\usepackage{hyperref}
\usepackage{fullpage}
\usepackage{makecell}
\usepackage{amsmath,amssymb,amsfonts}
\usepackage{algorithmic}
\usepackage{graphicx}
\usepackage{textcomp}
\usepackage{xcolor}
\usepackage{multirow}
\usepackage{braket}
\usepackage[caption=false]{subfig}

\newif\ifdraft
\ifdraft
\newcommand{\note}[1]{ {\textcolor{orange} { **Note: #1 }}}
\newcommand{\alnote}[1]{ {\textcolor{blue} { ***Andre: #1 }}}
\newcommand{\jknote}[1]{ {\textcolor{brown} { ***Johannes: #1 }}} 
\newcommand{\lhnote}[1]{ {\textcolor{orange} { ***Leo: #1 }}}
\newcommand{\fknote}[1]{ {\textcolor{orange} { ***Florian: #1 }}}
\newcommand{\jrfnote}[1]{ {\textcolor{pink} { ***Rudi: #1 }}}
\newcommand{\jrf}[1]{ {\textcolor{pink} { ***Rudi: #1 }}}
\newcommand{\crnote}[1]{ {\textcolor{teal} { ***Carlos: #1 }}}
\newcommand{\menote}[1]{{\textcolor{purple} {***Marvin: #1}}}
\newcommand{\lmnote}[1]{{\textcolor{green} {***Lukas: #1}}}
\newcommand{\cknote}[1]{{\textcolor{cyan} {***Chandan: #1}}}
\newcommand{\fsnote}[1]{{\textcolor{yellow} {***Francesc: #1}}}
\else
\newcommand{\note}[1]{ {\textcolor{orange} {}}}
\newcommand{\alnote}[1]{ {\textcolor{blue} {}}}
\newcommand{\jknote}[1]{ {\textcolor{brown} {}}} 
\newcommand{\lhnote}[1]{ {\textcolor{orange} {}}}
\newcommand{\fknote}[1]{ {\textcolor{orange} {}}}
\newcommand{\jrfnote}[1]{ {\textcolor{pink} {}}}
\newcommand{\jrf}[1]{ {\textcolor{pink} {}}}
\newcommand{\menote}[1]{ {\textcolor{purple} {}}}
\newcommand{\crnote}[1]{ {\textcolor{teal} {}}}
\newcommand{\lmnote}[1]{{\textcolor{green} {}}}
\newcommand{\cknote}[1]{{\textcolor{magenta} {}}}
\newcommand{\fsnote}[1]{{\textcolor{yellow} {}}}
\fi
\newcommand{\ignore}[1]{}

\begin{document}

\begin{frontmatter}

    \title{Quantum Computing for Automotive Applications}
    \author{Carlos A. Riofr\'io, Johannes Klepsch,  Jernej Rudi Finžgar, Florian Kiwit, Leonhard Hölscher, Marvin Erdmann, Lukas Müller, Chandan Kumar, Youssef Achari Berrada and Andre Luckow}
    
    \address{BMW Group Quantum Team, BMW Group, 80788 Munich, Germany}

    \begin{abstract}
        Quantum computing could impact various industries, with the automotive industry with many computational challenges, from optimizing supply chains and manufacturing to vehicle engineering, being particularly promising. This chapter investigates state-of-the-art quantum algorithms to enhance efficiency, accuracy, and scalability across the automotive value chain. We explore recent advances in quantum optimization, machine learning, and numerical and chemistry simulations, highlighting their potential and limitations. We identify and discuss key challenges in near-term and fault-tolerant algorithms and their practical use in industrial applications.  While quantum algorithms show potential in many application domains, current noisy intermediate-scale quantum hardware limits scale and, thus, business benefits. In the long term, fault-tolerant systems promise theoretical speedups; however, they also require further progress in hardware and software (e.\,g., related to error correction and data loading). We expect that with this progress, significant practical benefits will emerge eventually. 
        
    \end{abstract}
    
\end{frontmatter}

\section{Introduction}
\label{sec:intro}

Quantum computing promises to solve specific complex, industry-relevant computational problems more efficiently than classical computing~\cite{Bayerstadler2021}. The automotive industry, in particular, can benefit from quantum computing, with McKinsey estimating its economic impact up to \$3 billion by 2030~\cite{mckinsey2020quantum}. For example, quantum algorithms could be leveraged in vehicle engineering and manufacturing, from optimizing complex supply chains to enhancing materials for batteries and fuel cells~\cite{industry_reference_problems21}.  

The advancements in quantum hardware and algorithms are exemplified by the quantum supremacy experiment~\cite{Arute2019}, the progress towards quantum utility~\cite{Kim2023},  error correction improvements~\cite{acharya2024quantumerrorcorrectionsurface, reichardt2024demonstrationquantumcomputationerror}, and the integration of quantum-classical systems~\cite{alexeev2023quantumcentric}. This progress is the driver of the economic impact predicted by several studies, e.\,g., Hyperion~\cite{hyperion2020quantum}, McKinsey~\cite{mckinsey2020quantum}, MIT~\cite{HBR2022QuantumComputing}.

However, recently, the expectations have become more realistic. A 2024 BCG report~\cite{bcg2024quantum} continued to predict significant economic impact but highlights challenges, notably the limitations of Noisy Intermediate-Scale Quantum (NISQ) hardware~\cite{preskill2018} and the slow progress in quantum algorithms. Meanwhile, advances in classical computing -- especially in hardware like GPUs and AI algorithms -- have significantly raised the performance bar.

This chapter investigates theoretical and practical advancements and their impact on industrial applications. It provides a comprehensive overview of the current state of quantum algorithms. We address three key research questions: (i) For which automotive applications is there a prospective quantum advantage? (ii) What algorithms exist or will emerge for these applications in near-term, fault-tolerant, and quantum-inspired computing?  (iii) What are the key challenges and limitations in implementing these algorithms, and how can they be addressed?

This chapter is structured as follows: We begin by discussing the potential of quantum computing in the automotive industry in Section~\ref{sec:applications}. Our analysis then focuses on four problem domains: optimization, simulation, materials science, and machine learning (sections~\ref{sec:optimization}-\ref{sec:qml}). For each domain, we focus on three areas: (i) Near-term algorithms and techniques, such as variational quantum algorithms (VQAs)~\cite{vqa}, aiming to exploit today's quantum computers despite their limitations. (ii) Fault-tolerant quantum computing (FTQC) algorithms are expected to provide long-term quantum advantages based on proven theoretical advantages; however, they also demand more reliable and scalable quantum hardware. (iii) Quantum-inspired approaches which deliver measurable benefits today.  Section~\ref{sec:benchmarking} discusses the importance of benchmarks for evaluating quantum and classical solutions. We summarize our results and discuss future work in section~\ref{sec:future_direction}.

\section{Automotive Applications Areas}
\label{sec:applications}

The increasing complexity of the automotive industry has introduced significant computational challenges. A particular driver is automotive software, a critical enabler for connected and electric vehicles, advanced driver assistance systems, and user experiences and entertainment. Modern vehicles now contain over 100 million lines of code~\cite{bcg2022automotive}. This sophistication extends beyond vehicle design to processes like manufacturing, logistics, and sales, increasing computational demands across the entire value chain. Consequently, many optimization, simulation, and machine learning challenges have emerged. In the following, we focus on three main application areas: processes, material sciences, and engineering.

\emph{Processes:} Assembling vehicles is associated with many challenges related to the complexity of the product and the vast number of variants produced daily. Many optimization challenges arise in decision-making and planning, involving strategic, tactical, and operational decisions. Examples of such problems are robotic path optimization~\cite{Schuetz_2022}, vehicle configurations~\cite{Finzgar_2022}, route optimization, placement \& distribution problems~\cite{awasthi2023quantum, taqo}, line balancing, shift scheduling~\cite{krol2024qissquantumindustrialshift}, and vehicle sequencing~\cite{industry_reference_problems21}. Many problems fall into the category of combinatorial optimization and are often NP-hard when solved brute-force, making them challenging for traditional computing methods. Quantum computing offers the potential to handle the computational demands of these problems.

\emph{Materials Science:}  Computational quantum chemistry enables entirely new application areas, e.\,g., the exploration of materials science questions related to battery chemistry, lightweight materials, or fuel cell chemistry, enabling improvements to the vehicle's efficiency and performance. Classical computational methods such as Configuration Interaction (CI) are inherently limited in model size due to their computational complexity. Recent results of highly parallel implementations tackle up to 26 electrons in 23 orbital spaces~\cite{LargestFCI_JCTC_2024}. Computational methods such as Density Functional Theory (DFT) scale better with system size but lack a systematic accuracy improvement for calculating ground state energy for highly correlated systems. These interactions must be understood to accurately model the chemical interactions in battery cell materials or catalytic reactions in hydrogen fuel cells. Quantum computers promise more precise simulations of molecular properties and interactions~\cite{PhysRevResearch.4.023019}.

\emph{Engineering:} Vehicle engineering increasingly relies on computer-aided engineering (CAE) and numerical simulations to improve the vehicle's design, function, and quality while improving engineering efficiency. For example, the simulation of the aerodynamics of car bodies avoids costly and time-consuming experiments in wind tunnels. Further examples include crash, aero-acoustic,  cooling systems, or airborne noise optimization. These simulations can take days or weeks on large clusters and represent a significant part of the automotive industry's high-performance computing (HPC) workload. The demand is continuously increasing due to the development of new and more complex types of simulations, a wider variety of vehicle models and configurations, and new AI-based methods. 

Numerical simulations are used to understand physical processes related to the vehicle, e.\,g., the airflow around a vehicle, which is government by the Navier-Stokes equation, or the behavior of electrical systems described by Maxwell's equations. In practice, it is required to solve these equations over complex geometries under complex physical constraints. Quantum computing promises to solve these partial differential equations efficiently~\cite{Gaitan2020}.\\
\\
We will continue by providing concrete case studies that apply quantum computing in optimization (section~\ref{sec:optimization}), simulation (section~\ref{sec:simulation}), material science (section ~\ref{sec:material}), and machine learning (section~\ref{sec:qml}). We map the case studies to algorithms, which we categorize into near-term, fault-tolerant (FT), and quantum-inspired.

\section{Optimization}
\label{sec:optimization}

Combinatorial optimization plays a crucial role in various applications across science and industry. Typically, the challenge is to optimize (e.\,g., minimize the cost or duration of) a given task, such as the motion of robotic arms~\cite{DBLP:conf/kivs/MehtaMW17} or the scheduling of shifts~\cite{krol2024qissquantumindustrialshift} (cf. Tables~\ref{tab:applications} and~\ref{tab:case_studies_findings}). Optimizing such tasks can yield significant financial and qualitative benefits, especially as industries become increasingly complex. However, realizing such gains requires identifying a (near-)optimal solution within a combinatorially growing space of possible configurations. A large body of literature has been devoted to developing classical solvers, but they often struggle to find optimal solutions even for modestly-sized instances with hundreds of variables~\cite{Packebusch2016, puchinger}. Early promising theoretical (e.\,g., Grover~\cite{grover_1996}) and empirical results (e.\,g., in annealing~\cite{Hauke2020-tt}) sparked interest in quantum approaches to combinatorial optimization, both on near-term and fault-tolerant devices~\cite{abbas2023quantumoptimizationpotentialchallenges}.

\subsubsection*{Near-term Algorithms}

Near-term quantum optimization algorithms are classified into digital (gate-based) and analog methods. For gate-based methods, a large body of work has been dedicated to the quantum approximate optimization algorithm (QAOA)~\cite{farhi2014quantum} and its derivatives~\cite{Blekos2024}. QAOA is a hybrid quantum-classical algorithm comprising parameterized quantum gates optimized by classical optimizers. Recently, several performance limitations of QAOA have been identified~\cite{StilckFrana2021}, most notably due to the local nature of the ansatz~\cite{Bravyi2020-tz,Farhi2020-lw,Farhi2020-va}. Recursive QAOA (RQAOA)~\cite{Bravyi2020-tz, Bravyi2022-hyb} has been proposed to address the issue of locality through a recursive variable elimination strategy informed by correlations between variables obtained from QAOA circuits. 

\begin{figure}[t]
    \centering
    \includegraphics[]{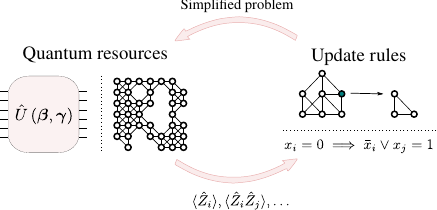}
    \caption{\textbf{QIRO:} A schematic representation of the quantum-informed recursive optimization (QIRO) algorithm. First, correlations between variables are obtained from quantum resources (e.\,g., QAOA or quantum annealers). The measured correlations are then used to simplify the problem using problem-specific update rules. The two steps are repeated until the problem is fully simplified or another termination criterion is met. 
    Figure adapted from~\cite{Finzgar2024qiro}.}
    \label{fig:qiro}
\end{figure}

Building upon RQAOA, a family of quantum-informed recursive optimization (QIRO) algorithms has been developed~\cite{Finzgar2024qiro} (see Figure~\ref{fig:qiro}). In QIRO, the optimization problem is simplified using problem-specific update rules informed by measurements on a quantum device. This approach leverages the user's understanding of a problem. For example, the update rules can enforce feasibility, which is crucial in real-world scenarios. Further, the performance of QIRO improves when the quality of the quantum information is enhanced, suggesting further performance gains with advancements in quantum hardware~\cite{Finzgar2024qiro, Fischer2024-pl}. 

Analog quantum optimization algorithms are typically referred to as quantum annealing~\cite{Hauke2020-tt}. In quantum annealing, the optimization problem is mapped to a quantum mechanical system, e.\,.g,  Rydberg atoms~\cite{Ebadi2022-nc} and superconducting flux qubits~\cite{Johnson2011}. The goal is to prepare a specific state (e.\,g., the ground state), which corresponds to a high-quality solution to the optimization problem. This is typically accomplished using different physical processes, most often using the adiabatic theorem of quantum mechanics~\cite{Born1928}. %
A notable recent example is solving the maximum independent set (MIS) problem on unit disk graphs by directly encoding it into the ground state of a Rydberg atom quantum device~\cite{pichler-mis,Ebadi2022-nc, Finzgar2024boqa}. 

Another challenge is mapping the optimization problem and the physical quantum system. Often, no direct mapping exists, and many mappings incur significant overheads~\cite{Knz2021}. Consequently, research focuses on developing these mappings, either focusing on specific hardware (e.\,g., neutral atom hardware~\cite{Nguyen2023-dy}), or particular problems (knapsack problem~\cite{awasthi2023quantum}, robot motion planning~\cite{Schuetz_2022, DBLP:conf/kivs/MehtaMW17}; see Table~\ref{tab:case_studies_findings}).

\subsubsection*{Fault-Tolerant Algorithms}
Most approaches in fault-tolerant quantum optimization algorithms leverage Grover's quantum search algorithm~\cite{grover_1996, gilliam2020optimizing}. An essential ingredient of Grover search is the so-called oracle, a function that identifies candidate solutions with desirable properties, e.\,g., feasibility. Krol et\,al.~\cite{krol2024qissquantumindustrialshift} construct an oracle for the shift scheduling problem, leading to a quadratic quantum speedup over a brute force search of the solution space. However, it remains unclear whether a quadratic speedup will be sufficient to achieve a quantum advantage in practice~\cite{Krol2023, Babbush2021-za}.

\subsubsection*{Quantum-Inspired Algorithms}
Several quantum-inspired optimization strategies have emerged as byproducts of the research in quantum optimization. The marriage of generative modeling and quantum-inspired tensor network methods resulted in the generator-enhanced optimization (GEO) framework~\cite{Alcazar2024-rk}. In GEO, a quantum-inspired generative model is used to propose new candidate solutions by identifying patterns in high-quality solutions. The performance of GEO has since been analyzed on the production planning use case, where it has been shown to perform on par with other conventionally used solvers~\cite{banner2023quantuminspiredoptimizationindustrial, vodovozova2025generativeenhancedoptimizationknapsackproblems}. 

The development of RQAOA and QIRO, also known as recursive shrinking algorithms, has led to further rounding heuristics for classical relaxation schemes. These algorithms take advantage of the fact that certain optimization problems can be relaxed into more tractable forms, such as linear or semi-definite programs, which allow for efficient solutions that can then be rounded to produce valid outcomes. Recursive freezing of variables, as employed in RQAOA and QIRO, has proven to be an effective technique for achieving this~\cite{Fischer2024-pl, wagner-ip}.

\subsubsection*{Discussion}
It is still unclear whether quantum computing can provide significant value for real-world optimization workloads~\cite{Sankar2024}. With large, noisy quantum devices, the key challenge is identifying hardware-native problem instances, e.\,g., maximum independent set problems for neutral atom hardware that can be embedded with minimal overhead while remaining difficult for classical solvers~\cite{Knz2021,Andrist2023,Wurtz:2022cqo}. While optimization of such problems on near-term quantum hardware devices has produced some impressive results~\cite{Ebadi2022-nc}, it is not yet feasible at the scales required for industrially relevant problems. However, with advancements in hardware and algorithms, scaling analyses suggest that quantum computing may eventually outperform classical methods for specific instances particularly suited to quantum devices~\cite{boulebnane2022, montanezbarrera2024}.

As we transition into the era of early fault-tolerant devices, identifying problems with structure that quantum computers can efficiently exploit becomes crucial for achieving greater-than-quadratic quantum speedups~\cite{aaronson2022structureneededhugequantum}, especially since Grover-type speedups alone are expected to be insufficient for practical quantum advantage~\cite{Babbush2021-za}. The quantum algorithms employed for such problems may incorporate or accelerate methods from the extensive literature on classical optimization~\cite{Sanders2020-jt, Montanaro2020, brandao2019, Ambainis2004}, or include non-unitary effects like measurements or engineered dissipation~\cite{Herman2023, Eder2024-kj}.

Finally, the algorithmic primitives developed primarily for near-term devices could provide useful building blocks for early fault-tolerant optimization algorithms. For example, a recently proposed heuristic strategy~\cite{Shaydulin2024-sz} uses the output of deep QAOA circuits as the initial state for amplitude amplification~\cite{Brassard2002}, a fault-tolerant routine that can be used for optimization. The results in Ref.~\cite{Shaydulin2024-sz} indicate a scaling advantage of this scheme when compared with state-of-the-art classical solvers for the considered problem.

\section{Simulation}
\label{sec:simulation}

\begin{figure}[t]
    \centering
    \includegraphics[]{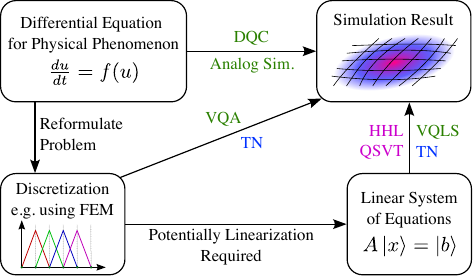}
    \caption{Overview of quantum algorithms for numerical simulations and the specific problems they solve. Near-term approaches are highlighted in green, fault-tolerant algorithms in pink, and quantum-inspired methods in blue. \alnote{can we add the color legend directly to the figure?}}
    \label{fig:overview_sim}
\end{figure}

In this section, we present an overview of the selected algorithms for three distinct subproblems arising in numerical simulations: (i) solving differential equations by determining an appropriate continuous function, (ii) finding a suitable discretized function, and (iii) solving linear systems of equations to obtain solutions on discretized space. \alnote{Can we label the arrows in Figure~\ref{fig:overview}?} Solving linear systems $Ax=b$ is a reoccurring routine in typical numerical simulation schemes, such as the finite difference, finite volume, or finite element method (FEM). Quantum algorithms can solve the analogous problem $A\ket{x}=\ket{b}$, commonly referred to as the quantum linear system problem (QLSP). Figure~\ref{fig:overview_sim} illustrates this classification and shows which problem each algorithm solves.

\subsubsection*{Near-term Algorithms}
Since the dynamics of quantum states follow the Schrödinger equation, a quantum computer inherently simulates that particular differential equation. Thus, a near-term quantum approach uses \emph{analog quantum simulation} and maps the desired differential equation onto the Schrödinger equation, a process referred to as Schrödingerization~\cite{Jin2024Analog}. After letting a quantum state evolve over time, it can be interpreted as a continuous simulation result. Until digital quantum computers are sufficiently advanced, this is a promising strategy for near-term devices. However, it is limited in its ability to handle nonlinear terms~\cite{PhysRevResearch.2.043102, Jin2023}, complex shapes, and boundary conditions relevant to industrial applications. 

Other near-term approaches fall in the category of VQAs, where quantum computers are used to evaluate a cost function, and conventional computers optimize the variational input parameters of these cost functions, similar to QAOA (see. section~\ref{sec:optimization}). Below, we briefly introduce some of the most promising and relevant VQAs, such as differential quantum circuits (DQCs)~\cite{KyriienkoPhysRevA.103.052416}, a VQA for nonlinear problems~\cite{LubaschPhysRevA.101.010301}, and the variational quantum linear solver (VQLS)~\cite{BravoPrieto2023}.

\emph{DQC:} This approach mirrors the concept of Physics-Informed Neural Networks (PINNs)~\cite{Raissi2019}. DQCs embed continuous functions described by differential equations directly into the cost function, enabling the use of automatic differentiation to represent function derivatives. This capability allows DQCs to address nonlinear problems effectively. Recent work studies the applicability of DQCs to solve the Navier-Stokes equations~\cite{JaderbergPhysRevA.109.042421} (cf. Table~\ref{tab:case_studies_findings}). A potential feature of DQCs could be that applications above the simulation layer could be incorporated into the training process (see PINNs for shape optimization~\cite{Zhang2024}).

\emph{VQAs for nonlinear problems:} A particular advantage of VQAs compared to fault-tolerant quantum algorithms is their ability to solve nonlinear problems. Many phenomena, such as fluid dynamics or structure mechanics, have nonlinear dynamics. By building a nonlinear cost function, VQAs have been proposed to solve the nonlinear Schrödinger equation~\cite{LubaschPhysRevA.101.010301} or the incompressible Navier-Stokes equations~\cite{Jaksch2023Variational}. In contrast to DQCs or analog quantum simulation, these algorithms solve differential equations on a discretized grid with finite difference operators.

\emph{VQLS:} The VQLS algorithm does not directly solve differential equations but instead addresses the QLSP, where $A$ is given as a linear combination of unitaries and the corresponding cost function penalizes a deviation between $A\ket{x}$ and $b$~\cite{BravoPrieto2023}. For instance, the VQLS method has been used to solve the advection-diffusion equation~\cite{Demirdjian2022}.

VQAs are generally interesting, as they can be studied on early quantum computers. Still, there are open questions about how well these methods perform at industrially relevant scales and how to overcome typical challenges such as barren plateaus~\cite{larocca2024reviewbarrenplateausvariational}.

\subsubsection*{Fault-Tolerant Algorithms}

The most prominent fault-tolerant algorithms for differential equations~\cite{Berry2014, Berry2017} solve the QLSP problem, e.\,g., the Harrow-Hassidim-Lloyd (HHL) algorithm~\cite{hhl, Childs2017} and the quantum singular value transform (QSVT)~\cite{MartynPRXQuantum.2.040203}. 

\emph{HHL:} The HHL algorithm uses quantum phase estimation (QPE)~\cite{nielsen_chuang_2010} to implicitly write the eigenvalues of matrix $A$ in a quantum register. It then stores the inverted eigenvalues in the quantum amplitudes, effectively inverting $A$. After reversing QPE and a postselection step, the algorithm prepares a state proportional to $\ket{x}$, given that the input state was $\ket{b}$. The overall complexity is $\mathcal{O}(\log(N)\kappa^2/\epsilon)$~\cite{hhl}, where $N$ corresponds to the dimension of $A$, $\kappa$ is its condition number quantifying the difficulty of inverting $A$, and $\epsilon$ is the precision of the solution. There is an exponential advantage in $N$ compared to the best-known classical methods. However, the promised speedup can be significantly reduced or even destroyed by considering the algorithmic overhead for state preparation, implementation of matrix oracles, or the readout of the full solution \cite{aaronson2015}. For example, in the case of FEM, the anticipated exponential speedup of the HHL algorithm is reduced to a polynomial speedup by considering the precision of the readout~\cite{MontanaroPhysRevA.93.032324}. \alnote{The HHL paragraph read very technically. Is there a way to simplify it and discuss the limitations?}\lhnote{I rearranged the sentences for better readability and moved the limitations up here.}

\emph{QSVT:} This approach applies a polynomial function to the singular values of a unitary matrix and is a generalization of the so-called quantum signal processing (QSP) method~\cite{Low2019hamiltonian}. \alnote{also here it is unclear for me what QSVT is and does? In which part of the process is it used to solve differential equations?} By block encoding the matrix $A$ in the upper left block of this unitary matrix, we can transform its singular values according to the chosen polynomial. If the polynomial is chosen to approximate $1/x$ up to the desired accuracy $\epsilon$, QSVT effectively inverts the matrix $A$. By applying the QSVT operator on $\ket{b}$, we again yield a state proportional to $\ket{x}$. The complexity of QSVT for this task is $\mathcal{O}(\kappa \log(\kappa/\epsilon))$~\cite{MartynPRXQuantum.2.040203}, which remarkably is independent of $N$. Here, the algorithmic complexity of block encoding and other subroutines is not accounted for, which can again quickly diminish the algorithmic advantage. Additionally, QSVT requires a classical precomputation of circuit parameters, which can become highly non-trivial. Hence, it is impossible to give a general statement about the utility of QSVT and HHL, since an efficient implementation depends on the structure, sparsity, and condition number of the linear system and, thus, on the concrete application.

One promising application based on QPE is the calculation of response functions of coupled oscillators, as studied in~\cite{danz2024calculatingresponsefunctionscoupled}. This application is particularly relevant to the industry as it could be used to reduce vehicle acoustic vibrations, for example. Since this algorithm only requires a product state as input, the state preparation can be executed efficiently and \alnote{Why can, in this case, state execution be executed efficiently?}\lhnote{I clarified it. The algorithm takes a product state as input which only requires n operations for n qubits. The product state is interpreted as a superposition of eigenstates but these details can be found in the reference.} the algorithmic complexity depends on the respective eigenvalues and desired accuracy. Another intriguing approach is to use the HHL algorithm as a subroutine within a larger algorithm to bypass the need for efficient state preparation or readout \cite{hhl}. This has been done, for example, by combining the QAOA and HHL algorithm~\cite{stein2023combiningqaoahhlalgorithm}. Specifically, the quantum state prepared by the quantum linear system solver can be used directly within the cost function of the optimization routine, avoiding a costly readout of the state. This approach can be applied to optimize the design of mechanical structures or systems~\cite{classiqClassiqCollaborates}.

\subsubsection*{Quantum-Inspired Algorithms}

Quantum-inspired methods~\cite{GarcaRipoll2021} for simulation mostly leverage classical tensor network (TN) techniques, originally developed for simulating quantum many-body systems, to solve non-quantum problems. Instead of approximating quantum states with tensor networks like Matrix Product States (MPS), one can approximate discretized functions and their derivatives using finite difference-based differential operators. 
The MPS format is particularly effective for functions with scale separations, utilizing a technique known as quantics encoding~\cite{Khoromskij2011, RitterPhysRevLett.132.056501}. This approach is relevant in fluid dynamics simulations, where turbulent flows exhibit significant scale separations~\cite{Gourianov2022quantum}. In particular, for 2D turbulence, it has been shown that the MPS format preserves the characteristic turbulent kinetic energy distribution~\cite{hoelscher2024quantuminspiredfluidsimulation2d} (cf. Table~\ref{tab:case_studies_findings}).

Quantum-inspired methods also offer insights for quantum algorithms, e.\,g.,  quantum-inspired tensor networks can be interpreted as quantum states or gates, and there exists a direct mapping from MPSs to a quantum circuit~\cite{rudolph2022decompositionmatrixproductstates}. Consequently, if data can be efficiently represented as a low-rank MPS, it can be efficiently loaded on a quantum computer~\cite{jobst2023efficient}. Moreover, TNs can be used to construct (nonlinear) cost functions of variational quantum circuits~\cite{LubaschPhysRevA.101.010301, Jaksch2023Variational} (see section \ref{sec:qml}). 

\subsubsection*{Discussion} 
A potential quantum advantage may emerge later than in other domains for numerical simulations. Near-term quantum approaches relying on VQAs face significant challenges, e.\,g.,  barren plateaus. Their heuristic nature makes it difficult to predict whether and for what problem structure an advantage emerges. This limits their immediate applicability to complex simulations. The DQC approach is still promising, as the model can include design parameters of simulation-based optimization methods, enabling efficient parameter space exploration. Analog quantum simulations show promise, especially as quantum hardware and Hamiltonian simulation techniques improve. However, applying these advancements to industrially relevant problems, such as encoding complex boundary conditions, remains challenging and uncertain. 

Even with perfect quantum computers, realizing a clear advantage in numerical simulations is not straightforward. The promised theoretical speedup by fault-tolerant algorithms, such as HHL or QSVT, cannot be directly transferred to typical numerical simulation workflows as long as additional required subroutines, e.\,g.,  state preparation, Hamiltonian simulation, block encoding, and information extraction, have a high complexity.

\section{Materials Science and Quantum Chemistry}
\label{sec:material}

The original idea of quantum computing was to simulate quantum systems using other quantum systems \cite{feynman1982simulating, Feynman+1988+523+548, doi:10.1126/science.273.5278.1073}. Modeling the properties of microscopic particles (e.\,g., electrons, atoms, or molecules)  can enhance our understanding and prediction of material behaviors, leading to advancements in applications such as drug discovery~\cite{Santagati2024, Yu2023}, battery design~\cite{IEEESpectrum_batteries, PhysRevResearch.4.023019}, and fuel cells~\cite{DiPaola:2023okw}.

Such applications are usually described by many interacting particles. As the complexity, i.\,e., memory and computation requirements, for simulating these many-body quantum systems scales exponentially with the number of particles, classical computers cannot keep up with the demands. Thus, it seems inevitable that a physical system with the same complexity scaling, i.\,e., a quantum system, could be used to simulate another. Simulating quantum systems' physical and chemical properties on a quantum computer presents one of the clearest paths to achieving quantum advantage, particularly in applications like materials science, quantum chemistry, and high-energy physics.

For material science and quantum chemistry, the starting point for quantum simulation is the Hamiltonian of the system. Usually denoted by $H$, the Hamiltonian describes the total energy of a given system, encodes the system's physical properties, and defines its evolution via, for example, the Schrödinger equation for a closed quantum system. In general, the physical systems are described via spin or molecular models. Spin models describe the system of the particles via spin-spin interactions. For example, the Heisenberg spin model~\cite{10.1088/978-0-7503-3879-0ch1} can describe the electromagnetic properties of materials~\cite{schollwöck2008quantum} and can readily be encoded into a quantum computer. 
Quantum systems can also be described by molecular models, which directly represent the electronic and nuclear interactions via electromagnetic forces among the atoms forming a molecule. This representation is commonly used in quantum chemistry. Ultimately, the molecular model can be mapped to a spin model via the second quantization and the Jordan-Wigner transformations~\cite{BRAVYI2002210}.

For many applications, the objective is to determine the eigenvalues (energies) and eigenstates of the model Hamiltonian, which is usually computationally expensive. In a quantum computer, however, once the Hamiltonian is written as a spin model, the procedure to find low-energy states requires the ability to measure the expected values of each of its terms with respect to a reasonably prepared trial state. A particularly important task of a quantum simulator is to determine the ground state of a given system, i.\,e., the state of minimal energy of the system. This application is relevant because many physical and chemical properties of a material or molecule depend on the properties of its ground state. Many algorithms can handle this task. Among the more prominent are the variational quantum eigensolver (VQE)~\cite{vqe} and variations of it, e.\,g., the generative quantum eigensolver (GQE) \cite{nakaji2024generative}, for near-term quantum devices; the quantum signal processing (QSP) and quantum-informed auxiliary field quantum Monte Carlo (QIAFQMC), for fault-tolerant quantum devices, which we briefly describe below and summarize in Table~\ref{tab:applications}.

\subsubsection*{Near-term Algorithms}
\fsnote{a NISQ (not an)}\alnote{fixed}

\emph{VQE:} This type of algorithm starts with a variational quantum circuit (VQC) ansatz, which attempts to prepare the system's ground state. In an iterative process, the variational parameters of the VQC are adapted via gradient descent, trying to minimize the system's energy \cite{vqa, vqe}. VQE is considered a near-term algorithm that is suitable for NISQ computers. However, it suffers from trainability issues when large systems are considered or as the depth and complexity of the VQCs grow; see the barren plateau problem \cite{larocca2024reviewbarrenplateausvariational}. When a suitable set of parameters for the VQC is found and a trial state is prepared, the expected value of the Hamiltonian can be estimated. Thus, its energy and eigenstate can be found.

\emph{GQE:} Different training methods have been developed to circumvent the VQE algorithm's trainability problems. One such example is to try to use classical neural networks to find the appropriate circuit ansatzes. It has been proposed to use a large language model (LLM) based on a pre-trained transformer architecture on relevant quantum circuits and their respective energy for a given Hamiltonian~\cite{nakaji2024generative}. The idea is to iteratively prompt the LLM to produce circuit ansatzes that reduce the system's energy while actively biasing the outcome probabilities towards lower energy states. This approach effectively transforms the continuous-variable optimization (for a fixed VQC ansatz) into a discrete combinatorial problem, which may alleviate standard VQE's trainability issues. 

Both VQA-based algorithms are suitable for near-term quantum devices. However, their practical implementation on current hardware remains limited. For example, Jattana et\,al.~\cite{PhysRevApplied.19.024047} simulate up to 40 qubits on classical hardware but perform a much more modest test using current quantum hardware. 

\emph{QIAFQMC: } More recently, there have been attempts to improve the performance of the quantum Monte Carlo algorithm for ground state finding, notoriously affected by computational limitations and the sign problem \cite{doi:10.1126/sciadv.abb8341}. Among these approaches, there have been attempts to enhance AFQMC methods \cite{https://doi.org/10.1002/wcms.1364} using a quantum computer \cite{Huggins2022, Qutac_QCAFQMC}. This approach uses a quantum computer to prepare suitable trial states that aid in the computation of the system's energy.

More complex, resource-intensive, and theoretically more effective algorithms exist for fault-tolerant quantum devices. We will describe a few of them below.

\subsubsection*{Fault-Tolerant Algorithms}

\emph{Qubitization or Block encoding:} All quantum circuits are unitary transformations. In this sense, unitary operators are native to quantum computers. There is no native way to encode arbitrary operators, e.\,g., general Hamiltonians or arbitrary operators, in a quantum register and operate over them. Such ability is necessary for quantum simulations. \crnote{We could cut the previous 4 sentences if we really want to without a lot of continuity loss, but I would advice against it. The sentences do explain why quantum simulators need Qubitization.} To simulate Hamiltonian dynamics using a quantum computer, it is necessary to use the technique of qubitization or block encoding \cite{Low2019hamiltonian}, which is an FT algorithm primitive. The technique consists of encoding an arbitrary operator inside a larger unitary operator through auxiliary qubits. This technique is probabilistic, which adds to the sample complexity of the algorithms. When this is achieved, additional algorithms can perform computations with the encoded data, as we comment below.

\emph{QSP:} This technique is a fault-tolerant algorithm and a critical tool for Hamiltonian simulations and ground state preparation tasks. QSP can apply a polynomial transformation to scalar quantities encoded in a quantum register~\cite{Low2019hamiltonian}. By using auxiliary qubits, QSP can implement complex polynomial transformations. This means that, in principle, any function that can be expanded via a Taylor series can be applied to the information encoded in a quantum computer.

For example, the quantum eigenvalue transformation (QEVT) algorithm uses QSP to approximate functions applied to the spectrum of the Hamiltonian of the system. First, qubitization is used to block-encode a spin-model Hamiltonian in a quantum register. Then, QSP is applied to compute a function of the encoded Hamiltonian. Finally, the system is measured, revealing its state and, thus, the result of the simulation. QEVT can be used to prepare the ground state of a Hamiltonian via imaginary-time evolution; see~\cite{Motta2020}. This technique aims to exponentially dampen all eigenvalues corresponding to energies higher than the ground state. Ideally, such transformation would prepare only the state with the lowest energy.

\subsubsection*{Discussion}

Quantum computing is especially suited to simulate relevant materials' physical and chemical properties. For the automotive industry, chemistry applications are the most prominent. They include simulation of the chemistry of lithium batteries and hydrogen fuel cells. For those purposes, VQC-based algorithms can be used in near-term quantum hardware. However, it is unclear whether they can handle workloads for industry-relevant scales while still producing useful results. Fault-tolerant algorithms, which are more resource-intensive, exist but are currently out of scope for near-term devices. These algorithms will be useful as less noise and better-controlled quantum hardware become available. Eventually, they may overtake classical algorithms when large-scale FT quantum devices exist.

\section{Machine Learning}
\label{sec:qml}

\fsnote{the integration OF quantum} \alnote{fixed}

Machine learning~\cite{Goodfellow-et-al-2016,bishop2023learning} and artificial intelligence (AI) have become critical for various industries~\cite{Singla2024StateAI}, including the automotive industry~\cite{8622357}, by enabling systems to identify patterns, learn from data, and support intelligent decisions. Generative AI systems support generating new, derived data but require large-scale machine learning models with billions of parameters. 

Progress in AI is driven by the scaling laws that describe the improvement of AI as a function of the progress in computing, data, and model parameters~\cite{kaplan2020scalinglawsneurallanguage}. Similarly, Sutton's "bitter lesson"~\cite{sutton2019bitter} emphasizes the long-term success of algorithms that leverage increased computational power over those relying on complex, domain-specific optimizations. As classical computing infrastructures hit scaling limits~\cite{semianalysis2024}, quantum computing may eventually provide the computational capabilities needed to align with this principle.

Quantum machine learning (QML)~\cite{Biamonte_2017,Schuld_2014} integrates quantum algorithms with machine learning, aiming to reduce training times and enhance the scalability and quality of models. While some QML algorithms promise theoretical improvements in computational complexity~\cite{Sweke2021quantumversus,Liu_2021,Huang_2021,PhysRevA.107.042416}, the practical advantages of quantum machine learning have yet to be conclusively established on large-scale industry-relevant datasets. In the following sections, we describe near-term, fault-tolerant, and quantum-inspired QML, concluding with a discussion on the limitations of QML.

\subsubsection*{Near-term Algorithms}
\fsnote{Missing word between notably and leverage?} \alnote{fixed}

\emph{Representing classical data on a quantum computer:}\
QML algorithms can be applied to both quantum and classical data. For classical data, input data must be mapped to a quantum state using a procedure referred to as data or feature encoding. Notably, encoding methods that leverage entanglement, such as amplitude encoding, are particularly interesting because they allow for exponential data compression by encoding a feature vector of size $N$ into $\log N$ qubits. However, methods like amplitude encoding incur an exponential runtime in the number of qubits, and any potential quantum speed-up is lost at the encoding stage~\cite{aaronson2015}. A promising solution lies in using approximations to address the challenge of exponential runtime in amplitude encoding. Jobst et al.~\cite{jobst2023efficient} show that classical data with an algebraically decaying Fourier spectrum can be well-approximated by a quantum circuit with a linear number of nearest-neighbor two-qubit gates.

\begin{figure*}[t]
    \centering
    \includegraphics[width=\linewidth]{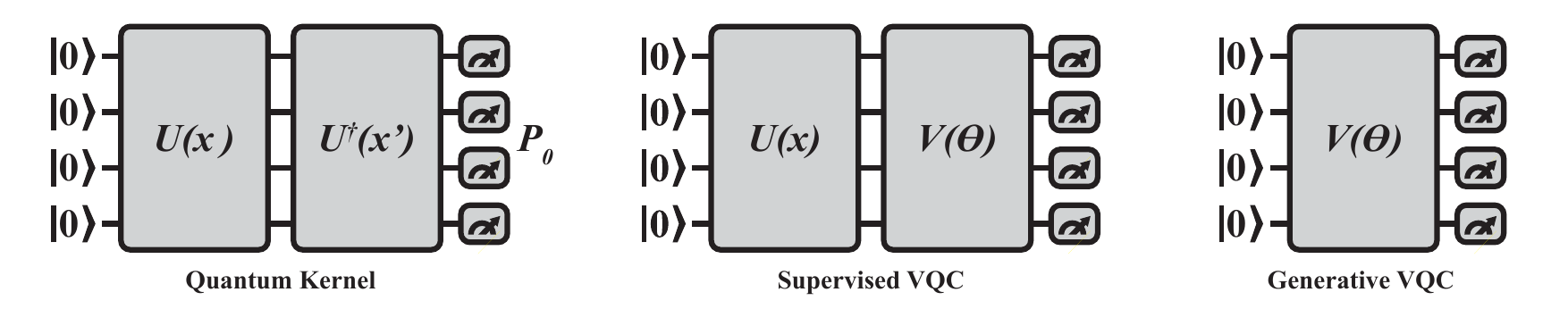}
    \caption{Illustration of different types of quantum circuits used in quantum machine learning. The quantum kernel model (left) utilizes a unitary operation \( U(x) \) followed by its Hermitian conjugate \( U^\dagger(x') \) to calculate the inner product between two feature vectors. The supervised VQC model (middle) combines data encoding with a parameterized quantum circuit \( V(\theta) \), allowing for the optimization of parameters \( \theta \) during supervised learning tasks. The generative VQC (right) consists solely of a parameterized quantum circuit \( V(\theta) \) designed to generate quantum states that can model complex data distributions.}
    \label{fig:model_types}
\end{figure*}

QML encompasses various methods, including quantum kernel techniques and variational models. The latter can be applied to both supervised and generative tasks. Figure~\ref{fig:model_types} illustrates these different models, which will be discussed in the following.

\emph{Quantum kernel methods:}\
Quantum kernel methods~\cite{Schuld_2019b,Havl_ek_2019} are one approach in QML for supervised learning tasks. They leverage quantum computing by replacing the classical kernel of a support vector machine (SVM) with a quantum kernel. In this approach, a classical data vector $\mathbf{x}$ is mapped into a high-dimensional feature space using a feature map that is represented by a parameterized quantum circuit $U(\mathbf{x})$. Quantum computers can efficiently evaluate the inner products between data points in this quantum feature space. By determining all pair-wise inner products of the embedded data points, the kernel matrix is generated. This kernel matrix is then used within the SVM framework to perform classification or regression tasks. 

\emph{Supervised training of VQCs:}\
Supervised training of VQCs~\cite{PhysRevA.98.032309,farhi2018classificationquantumneuralnetworks,Schuld_2020} involves optimizing the parameters of a VQC to minimize a predefined cost function, typically derived from labeled data. The circuit processes input data $x$ encoded into quantum states by a feature map $U(x)$. This state is then evolved via a VQC $V(\theta)$, before the measurement is performed. The expectation value of the observable acts as the prediction of the model. The parameters $\theta$ are tuned during the training to minimize a cost function. A method for computing gradients on a quantum computer is the parameter-shift rule~\cite{PhysRevA.98.032309,Schuld_2019}.

\begin{figure*}[t]
    \centering
    \includegraphics[width=0.9\linewidth]{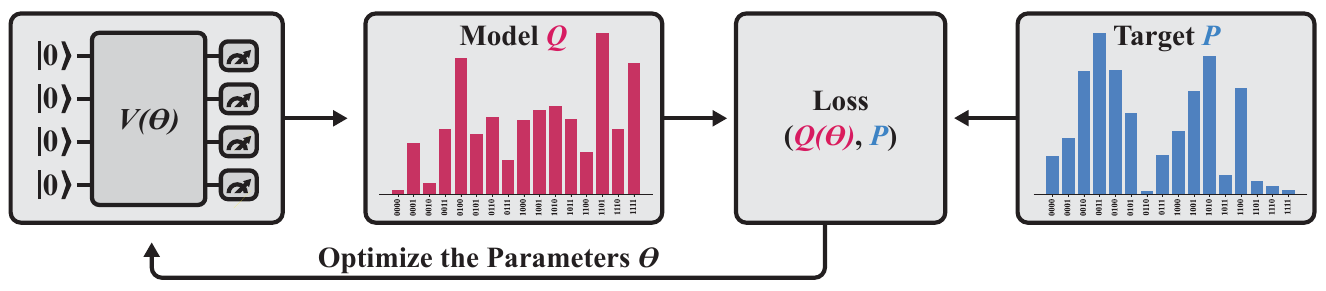}
    \caption{Training procedure of a quantum generative model. The model $V(\theta)$ is parameterized by a set of parameters $\theta$ and is initialized with all qubits in the \(|0\rangle\) state. The quantum circuit $V(\theta)$ generates a probability distribution $Q(\theta)$ over possible measurement outcomes, represented by binary strings. The goal is to optimize the parameters $\theta$ such that the generated distribution $Q(\theta)$ closely matches the target distribution $P$. The loss function $\text{Loss}(Q(\theta), P)$ quantifies the difference between the generated and target distributions, guiding the optimization of $\theta$.}
    \label{fig:generative}
\end{figure*}

\emph{Generative training of VQCs:}\
Quantum states prepared by VQCs naturally represent probability distributions that can be used to generate synthetic data. Therefore, VQCs can also be used for generative learning. Instead of fitting the model's output to the data label, the output is fitted to a distribution. Two widely used training routines in quantum generative modeling, as depicted in Figure~\ref{fig:model_types}, are the quantum circuit Born machine (QCBM)~\cite{benedetti2019generative,liu2018} and the quantum generative adversarial network (QGAN)~\cite{PhysRevA.98.012324,Lloyd_2018}, as depicted in Figure~\ref{fig:generative}. For a comprehensive review of these methods, see the referenced literature~\cite{10.1145/3655027}. The QCBM aims to minimize a statistical distance such as the Kullback-Leibler (KL) divergence, between the probability distributions of the generated and target data by adjusting model parameters with gradient-based or gradient-free optimization techniques.

In contrast, the QGAN adopts a framework similar to classical generative adversarial networks~\cite{goodfellow2014generative}, which operate on the principle of adversarial training through a minimax game involving two neural networks: the generator and the discriminator. The generator produces synthetic data, while the discriminator distinguishes between real and generated samples. In QGANs, classical components are replaced with a VQC. The parameters are typically updated using gradient descent, with gradients calculated, e.\,g., via the parameter-shift rule. The classical components in a QGAN are optimized using a classical optimizer, with gradients computed through backpropagation.

\subsubsection*{Fault-tolerant Algorithms}

Finding quantum advantage requires both the engineering task of building a quantum computer and finding problems where a quantum algorithm provides a superpolynomial speedup compared to the best-known or classical algorithms~\cite{preskill2012quantumcomputingentanglementfrontier}. Estimating the quantum advantage threshold is inherently tied to the performance of the most efficient classical algorithms available; as these classical algorithms improve, the threshold for quantum advantage shifts. Notably, some quantum machine algorithms initially believed to offer super polynomial speedups, such as quantum recommendation systems~\cite{kerenidis2016quantumrecommendationsystems} or quantum Principal Component Analysis (PCA)~\cite{Lloyd_2014}, were later shown to lack this advantage when more efficient classical alternatives were proposed~\cite{PhysRevLett.127.060503}.

Most fault-tolerant quantum machine learning algorithms, focus on accelerating the linear algebra, e.\,g., using HHL~\cite{hhl} or the quantum least-square fitting algorithm~\cite{PhysRevLett.109.050505} (see~\cite{dalzell2023quantum}, section 9, for survey). Liu et al.~\cite{Liu_2024} propose a quantum algorithm for improving the efficiency of training large-scale machine-learning models, particularly in the context of stochastic gradient descent.\alnote{some more details on how sgd is improved would be useful} Johri et al.~\cite{Johri2020NearestCC} propose an algorithm with an exponential speedup for computing inner products that can be used to speed up the convolutions in convolutional neural networks.

Another example of fault-tolerant QML is Bayesian inference~\cite{Low_2014}. Exact Bayesian inference is $\#P$-hard, and classical algorithms perform approximate inference, which remains NP-hard. However, quantum rejection sampling~\cite{Ozols_2012} offers a way to accelerate Bayesian inference. The Bayesian network can be represented as a quantum circuit in this approach, with the conditional probabilities encoded as controlled quantum operations.

\subsubsection*{Quantum-Inspired Algorithms}
Quantum-inspired approaches have been proposed in both supervised~\cite{Reyes_2021,Stoudenmire_2018,NIPS2016_5314b967,Huggins_2019,TN_Xanadu} and unsupervised~\cite{PhysRevX.8.031012,Cheng_2019} machine learning. The core idea behind these quantum-inspired methods is to take a tensor network (TN)~\cite{Or_s_2014,Or_s_2019}, such as matrix product states (MPS) or tree tensor networks (TTN), and fit it to the given data. TNs are mathematical structures that efficiently represent high-dimensional data and can be adapted to learn from datasets by optimizing their parameters. The structure of the TN enables the use of algorithms like density matrix renormalization group (DMRG)~\cite{PhysRevLett.69.2863} to compute the gradients of the model efficiently. One advantage of machine learning with tensor networks is that the complexity of the model (i.\,e., the number of parameters) can be dynamically adapted during training. Another application of tensor networks is image reconstruction~\cite{PhysRevX.8.031012}.

\subsubsection*{Discussion}

QML, like other VQAs, is hindered by training difficulties and hardware restrictions, which limit model complexities and the number and quality of measurements. In the following, we will explore key challenges.

\emph{Data Loading:} Amplitude encoding enables to represent \(N\) classical features in \(n = \log N\) qubits, offering an exponentially compact representation. \alnote{but state preparation is still exponentially complex?} While QML algorithms that scale polynomially with qubits promise logarithmic runtime dependency on dataset size, the exponential cost of preparing amplitude-encoded states often negates these speedups, raising the question of whether algorithms can be developed to prepare these states with a run time polynomial in $n$.

\emph{Barren plateaus~\cite{larocca2024reviewbarrenplateausvariational,McClean_2018, cerezo2024doesprovableabsencebarren}:} In quantum computing, Barren plateaus occur when the gradient of the cost function becomes nearly zero over large regions of the parameter space, making training quantum algorithms extremely costly. This happens when the variance of the loss function's partial derivatives decays exponentially with the number of qubits, creating a flat loss landscape that makes finding a minimum exponentially difficult. Cerezo et al.~\cite{cerezo2024doesprovableabsencebarren} argue that variational quantum circuits designed to avoid barren plateaus \alnote{What are the criteria for such a circuit?} may be classically simulable, potentially diminishing a quantum advantage.

\emph{Backpropagation:} Machine learning heavily relies on backpropagation for training neural networks at scale.
Backpropagation~\cite{rumelhart_learning_1986} efficiently calculates gradients by reusing intermediate computations, incurring a cost roughly equivalent to running the function. In contrast, no such efficient gradient computation exists for VQCs~\cite{NEURIPS2023_8c3caae2}. Parameter-shift rules add overhead proportional to the number of parameters, posing a key challenge for scaling quantum machine learning (QML) approaches.

\emph{Noise:} Noise in quantum systems presents a significant challenge for QML, particularly when training variational quantum circuits. Quantum noise arises from various sources, including decoherence, gate errors, and measurement errors, which degrade the fidelity of quantum states and operations~\cite{preskill2018}. This degradation can lead to an inaccurate estimation of cost functions and gradients, exacerbating the barren plateau phenomenon by introducing additional uncertainty into the optimization process~\cite{Wang_2021}. Furthermore, noise increases the difficulty of reliably executing quantum circuits, necessitating error mitigation strategies that often come with a substantial resource overhead. Compilation techniques~\cite{9180791} are essential for mitigating noise in quantum systems. These methods optimize circuits by reducing their depth, minimizing error-prone operations, and strategically placing gates to avoid noisy qubits. Application-aware approaches~\cite{Quetschlich:2024oui} tailor circuits to a figure-of-merit defined by a specific quantum algorithm.

Fault-tolerant QML algorithms, while theoretically promising, struggle with different practical barriers, e.\,g., data loading and the necessity for quantum RAM (QRAM) that can utilize superposition to facilitate the efficient retrieval of training data. 

The next section discussed the current state of quantum application benchmarks.

\section{Benchmarking}
\label{sec:benchmarking}

Benchmarking is important for understanding the capabilities of current and emerging quantum computing hardware, software, and algorithms. Unlike classical computation benchmarks, like LinPack~\cite{LinPack} or SPEC~\cite{SPEC}, there is not yet a common understanding of standardized metrics. Quantum computing benchmarking can be classified into three categories: system, aggregate, and application-level benchmarks~\cite{e24101467}. System benchmarks focus on the basic physical properties of qubits and quantum gates on the quantum hardware level. Examples are T1 and T2 relaxation times, gate fidelity, and error rates (see Eisert et\,al.~\cite{Eisert2020} for an overview). 

Aggregate benchmarks evaluate a larger part of the quantum computing stack, considering a broader range of attributes. An example is Quantum Volume (QV), which considers the number of qubits, the quality of gate operations, and the accuracy of measurements~\cite{cross2019validating}. While the QV is a well-established benchmark, it has known drawbacks like its strong dependency on high-quality qubits, which favors technologies like ion-trapped quantum computers (which reach QV values of over a million~\cite{quantinuum_2024, IonQ_QV}) compared to superconducting hardware (currently reaching QV values well below 1000~\cite{IBM_QV}). Hence, other aggregate benchmarks like circuit layer operations per second (CLOPS)~\cite{wack2021qualityspeedscalekey} or generalized volumetric benchmarks~\cite{Blume_Kohout_2020} based on QV address these shortcomings and offer alternatives.

Application-level benchmarking focuses on the end-to-end usability of quantum computers' overall hardware, software, and algorithm stack levels, evaluated for individual use cases. The benefit of such benchmarks is the direct comparison of relevant quantum technologies, models, algorithms, and (hyper)parameters for use case owners to decide the best fit for their application. Bowles et al. provide QML benchmarks~\cite{bowles2024betterclassicalsubtleart,qml_benchmarks} with tools to compare the performance of near-term QML and classical machine learning models on supervised learning tasks. Another example of an application-level metric is Q-Score~\cite{Martiel_2021}. Q-Score measures the effective number of qubits to solve a well-known NP-hard optimization problem, the Max-Cut problem, with QAOA. The Q-Score's scope is actively extended to other relevant applications like Maximum Cardinality Matching~\cite{barbaresco2024bacqapplicationorientedbenchmarks}.

Several quantum computing benchmarking suites and frameworks have been developed to facilitate application-level benchmarking. These frameworks offer users and developers a way to run benchmarks and implement benchmarking workflows while not (necessarily) being equivalent to a metric. Instead, they often provide several performance measures, which can be generic (like computation times) or application-specific. Typically, application-specific metrics reflect the solution quality, like the distance traveled in a traveling salesperson problem or the Kullback-Leibler divergence of synthetic datasets of generative models~\cite{kiwit2024benchmarking}. Well-defined metrics in benchmarking suites must capture the trade-off between system performance and the solution quality between multiple quantum technologies. %

An example of an application-centric benchmark suite is the Quantum Economic Development Consortium (QED-C):  Application-Oriented Performance Benchmarks for Quantum Computing. Released as an open-source suite in 2021~\cite{Lubinski_etal_2021}, it initially focused on simple subroutines and applications like Grover's Search, Monte Carlo Sampling, and Shor's Period Finding, using normalized circuit fidelity as the primary metric. It has since expanded to include 15 applications, along with additional metrics and features~\cite{Lubinski_etal_2023,lubinski2024quantum}. It supports a variety of quantum libraries and hardware and is considered the most mature application-oriented benchmarking framework available.

Alternatives to the QED-C suite offer advantages in certain aspects of quantum computing benchmarking. The Quantum Computing Application Benchmarking (QUARK) framework is an open-source project that strongly focuses on modularity and flexibility to facilitate the integration of new application-level benchmarks comprised of datasets, algorithms, and metrics by developers from the community~\cite{QUARK, Finzgar_2022}. QUARK currently provides benchmarks for six applications, including optimization use cases like the maximum independent set problem and the auto-carrier loading problem, as well as a quantum machine learning application with generative modeling~\cite{kiwit2023application}.

MQT Bench is a benchmarking suite developed as part of the Munich Quantum Toolkit (MQT), which emphasizes the benchmarking of quantum algorithms at the circuit level~\cite{Quetschlich_2023}. With over 70,000 benchmark circuits, MQT Bench offers one of the world's largest quantum circuit libraries. Another algorithm-level benchmarking framework is SupermarQ, which aims to apply classical benchmarking methods systematically to quantum computing~\cite{tomesh2022supermarqscalablequantumbenchmark}. Similar to classical benchmark suites, like LinPack~\cite{LinPack}, PARSEC~\cite{PARSEC} or MLPerf~\cite{MLPerf}, SupermarQ aims to include a variety of applications (like Bell inequality tests and Hamiltonian simulation) and benchmarks state-of-art technology (e.\,g., IBM, IonQ, and AQT) to provide relevant benchmarks for a large number of use cases. QPack evaluates the performance of quantum hardware and simulators for QAOA and VQE applications in four categories (capacity, scalability, accuracy, and runtime) based on relevant execution data~\cite{donkers2022qpackscoresquantitativeperformance}.

This overview of quantum computing benchmarks highlights the field's diversity. While multiple perspectives help identify metrics, they can also complicate fair evaluations~\cite{proctor2024benchmarkingquantumcomputers}, with results varying across suites and metrics sometimes favoring specific providers~\cite{dehghani2021benchmarklottery}. Benchmarking often emphasizes quantum advantages without fair comparisons to classical alternatives, as noted in quantum machine learning~\cite{bowles2024betterclassicalsubtleart}. The community is moving toward greater collaboration and standardization to address these issues, focusing on developing real-world performance measurements and user-friendly benchmarking workflows~\cite{IEEE_Benchmarking, DIN_Spec}.

\section{Conclusion and Future Directions}
\label{sec:future_direction}

\crnote{we should make the following sections parallel: We either add a subsection name to the first paragraphs, to match 'benchmarking', 'resource estimation', etc. or we erase those subsections names. Let's try adding 'general'}\alnote{ok}

\begin{table*}[t]
    \footnotesize
\centering
    \begin{tabular}{|>{\raggedright\arraybackslash}p{2.1cm}|>{\raggedright\arraybackslash}p{3.3cm}|>{\raggedright\arraybackslash}p{2.cm}|>{\raggedright\arraybackslash}p{2.2cm}|>{\raggedright\arraybackslash}p{4.6cm}|}
        \hline
        \textbf{Problem Domain} & \textbf{Applications} & \textbf{Algorithms} & \textbf{Algorithm Classes} & \textbf{Case Studies} \\ \hline

        Optimization &\multirow{3}{3.3cm}{Robot motion, vehicle configuration, shift scheduling, warehouse capacity}
                                                            &Near-term   &QIRO~\cite{Finzgar2024qiro}, quantum annealing~\cite{Finzgar2024boqa}         & Robot motion~\cite{DBLP:conf/kivs/MehtaMW17,Schuetz_2022},\newline supply chain/knapsack~\cite{awasthi2023quantum}, vehicle configuration/SAT~\cite{Finzgar_2022} \\ \cline{3-5} 
                                                            & &Fault-tolerant         &Grover &Shift scheduling~\cite{krol2024qissquantumindustrialshift}  \\ \cline{3-5}                                                                                                                        
                                                            & &Quantum-inspired &GEO~\cite{Alcazar2024-rk}, 
                                                            shrinking algorithms~\cite{Fischer2024-pl}  &Production planning~\cite{banner2023quantuminspiredoptimizationindustrial} \\ \hline \hline

        Quantum Machine Learning &\multirow{3}{3.3cm}{Classification, synthetic data generation, data loading, quality control, fraud detection}  
            &Near-term  & VQA &Data generation with QGAN/QCBM~\cite{10.1145/3655027}, image classification~\cite{shen2024}, image compression~\cite{jobst2023efficient} \\ \cline{3-5}
            
            & &Fault-tolerant &HHL & PCA~\cite{Lloyd_2014}, stochastic gradient descent~\cite{Liu_2024}, Bayesian inference~\cite{Low_2014} \\ \cline{3-5}
            
            &        &Quantum-inspired & Tensor networks &Image classification~\cite{TN_Xanadu} \\ \hline \hline
        
        Numerical Simulation &\multirow{3}{3.3cm}{Computational Fluid Dynamics (CFD), structure mechanics, acoustics, thermodynamics}
                                    &Near-term &DQC~\cite{KyriienkoPhysRevA.103.052416}, VQLS~\cite{BravoPrieto2023}, VQA~\cite{LubaschPhysRevA.101.010301}, analog simulation~\cite{Jin2024Analog} & Flow past cylinder~\cite{JaderbergPhysRevA.109.042421}, advection-diffusion equation~\cite{Demirdjian2022}, Navier-Stokes equations~\cite{Jaksch2023Variational}, nonlinear Schrödinger equation~\cite{LubaschPhysRevA.101.010301}\\ \cline{3-5} 
                       &            &Fault-tolerant &HHL~\cite{hhl}, QSVT~\cite{MartynPRXQuantum.2.040203}, QPE~\cite{nielsen_chuang_2010} & Unit commitment problem~\cite{stein2023combiningqaoahhlalgorithm}, FEM~\cite{MontanaroPhysRevA.93.032324} \\ \cline{3-5} 
                       &                    &Quantum-inspired &Tensor networks & Turbulent CFD \cite{Gourianov2022quantum,hoelscher2024quantuminspiredfluidsimulation2d}, Navier-Stokes equations~\cite{kornev2023numericalsolutionincompressiblenavierstokes, Peddinti2024Quantum, KiffnerPhysRevFluids.8.124101} \\ \hline \hline

        Materials Science and Quantum Chemistry & \multirow{2}{3.3cm}{Simulation of physical properties of materials, e.\,g., drug discovery, battery and fuel cell chemistry}
                        &Near-term &VQE \cite{vqe}, GQE \cite{nakaji2024generative}, QMC~\cite{Montanaro_2015}, QIAFQMC \cite{Huggins2022}   & Ground state preparation, time evolution\\ \cline{3-5} 
                    &   &Fault-tolerant &Quantum signal processing \cite{Low2019hamiltonian}
                    &Ground state preparation\\ \hline

    \end{tabular}
    \caption{Summary of problem domains, applications, and algorithms: This table groups the conducted case studies by the algorithms used (including near-term, fault-tolerant, and quantum-inspired methods) and applications.
     \label{tab:applications}}
\end{table*}

\emph{General:} This chapter examined the current state of quantum algorithms and their applications in the automotive industry. 
Table~\ref{tab:applications} summarizes the problem domains, applications, algorithms and case studies conducted using these algorithms. We categorize the algorithms into near-term, fault-tolerant, and quantum-inspired methods. While much research has been focused near-term algorithms and application, increasingly fault-tolerant methods are investigated.

\begin{table*}[t]
    \footnotesize
    \centering
        \begin{tabular}{|>{\raggedright\arraybackslash}p{1.5cm}|>{\raggedright\arraybackslash}p{2.7cm}|>{\raggedright\arraybackslash}p{4.5cm}|>{\raggedright\arraybackslash}p{6cm}|}
            \hline
            \textbf{Problem Domain} & \textbf{Case Studies} & \textbf{Used Hardware} & \textbf{Problem Complexity} \\ \hline
    
            Optimiz\-ation & Robot motion~\cite{Finzgar_2022, Schuetz_2022} & Quantum Annealing with D-Wave 2000Q (2,048 qubits) and D-Wave Advantage 4.1 (5,000 qubits).\newline D-Wave qbsolv (hybrid quantum-classical). &2000Q: 2 seams of robot motion problem with constraints or 8 nodes in traveling salesperson graph (2000Q) \newline Advantage 4.1: 3 seams of robot motion problem with constraints and 14 nodes of traveling salesperson problem \newline 
            qbsolv: up to 30 seams of robot motion problem \newline
            classical: up to 71 seams of robot motion problem. \\ \cline{2-4}
            
            & Knapsack~\cite{awasthi2023quantum} & QAOA, VQE with up to 19 simulated qubits, annealing with D-Wave 2000Q (2,048 qubits) and  Advantage 6.1 (5,760 qubits). &Knapsack with two knapsacks and six items on 19 qubits \\ \cline{2-4}
            
            & Shift Scheduling~\cite{krol2024qissquantumindustrialshift} &Qiskit tensor network simulator with up to 48 qubits. &Simplified model with a body shop and a paint shop that share a buffer for up to 2 days. \\ \cline{2-4}

            & Maximum Independent Set~\cite{Finzgar2024qiro, Finzgar2024boqa} &QuEra Aquila with 137 qubits. &Network design optimization with up to 137 nodes. \\ \cline{2-4}

            & Vehicle configuration~\cite{Finzgar_2022} & Quantum Annealing with D-Wave 2000Q (2048 qubits). & Up to 100 possible features for a single vehicle.\\
            \hline
    
            QML 
            & Data generation with QGAN/QCBM~\cite{10.1145/3655027} & State vector simulation with Qiskit/Pennylane and Jax, IonQ harmony with 8 qubits~\cite{kiwit2023application}, Qiskit noise simulation with up to 12 qubits~\cite{kiwit2024benchmarking}. &2-3 dimensional data with up to 50,000 to 100,000 item datasets on 8-12 qubits  \\ \cline{2-4}
            
            & Image classification~\cite{shen2024} &IBM Quantum Kolkata with up to 11 qubits.  &Fashion MNIST comprising 70,000 labeled grayscale square images. \\ \hline
            
            Numerical Simulation & DQC simulation of 2D flow~\cite{JaderbergPhysRevA.109.042421}   &Unspecified hardware for PennyLane simulation.  &Flow past a 2D cylinder with a Reynolds number of $\mathrm{Re}=100$ with 6 qubits.\\ \cline{2-4}
                        
            & TN simulation of 2D turbulence~\cite{hoelscher2024quantuminspiredfluidsimulation2d} &NVIDIA H100 GPU with 80 GB of memory. & Two periodic turbulent flows with Reynolds numbers up to $\mathrm{Re} = 10^7$. The MPSs representing velocities would correspond to 52 qubits with limited entanglement. 
            \\ \hline
            
            Quantum Chemistry & Fuel Cell~\cite{DiPaola:2023okw} &Quantinuum H1 using 6 qubits & Simulation of Oxygen Reduction Reaction (ORR) on Platinum surface with active space of 2 electrons in 3 orbitals. 
            \\ \hline

        \end{tabular}
        \caption{Case Studies and Findings in Quantum Computing Applications: While quantum algorithms show advances in selected metrics, the scale of these studies is far away from business relevance.}
        \label{tab:case_studies_findings}
\end{table*}

\emph{Quantum Hardware and Simulators:} 
The utilized hardware platforms range from D-Wave's quantum annealers to Quantinuum and IonQ's quantum processors and classical simulators. Additionally, the table captures the scale and complexity of the problems addressed, offering insight into quantum algorithms' current capabilities and limitations as they are applied to real-world scenarios. In the following, we will discuss the key takeaways per problem domain, followed by the overarching trends.

Quantum hardware has significantly progressed. However, this progress has not matched initial industry expectations, particularly in scale. This limitation is evident from the data presented in Table~\ref{tab:case_studies_findings}, which summarizes our recent case studies. Our case studies utilized various quantum hardware platforms, e.\,g., approximately 5,000 qubits on D-Wave's Advantage 4.1, 137 qubits on QuEra's Aquilla, 11 qubits on IBM's Kolkata,  6-10 qubits on Quantinuum's H1 and IonQ's Harmony machines.

Classical state vector simulations were used in all case studies and provided an important baseline for quantum hardware experiments.  For selected QML experiments, we utilized GPU simulations~\cite{bayraktar2023cuquantum}, observing significant speedups of up to 300~\cite{kiwit2023application}. Further, approximate tensor network-based simulations help investigate larger quantum systems, e.\,g., with approximately 50 qubits for the optimization~\cite{krol2024qissquantumindustrialshift} and turbulence simulations~\cite{hoelscher2024quantuminspiredfluidsimulation2d} case studies.

The limitations of today's noisy quantum hardware significantly constrain near-term quantum algorithms across all application areas and problem domains. For example, all VQAs are subject to trainability and scalability (e.\,g., due to barren plateaus). While advancements in hardware may alleviate some of these issues, it remains uncertain whether these improvements will suffice to achieve a practical quantum advantage. VQA studies often compare quantum and classical machine learning on a small scale in carefully selected settings. Despite theoretical speedups, fault-tolerant methods often only address specific parts of algorithms and applications. Further, further challenges that need to be solved, e.\,g., data loading and extraction need to be solved while minimizing the number of measurements required~\cite{xanadu_qml_advantage}.

Despite the current practical limitations, quantum computing is progressing in all critical aspects, e.\,g.,  the number of physical qubits has doubled every one to two years since 2018~\cite{bcg2024quantum} and the quality of these qubits improved significantly (e.\,g., Quantinuum H2 system with 99.99\,\% single qubit fidelity~\cite{quantinuum_h2}). In addition, multiple qubit modalities, i.\,e., superconducting qubits, trapped ions, neutral atoms, and photonics, are advancing concerning scale and quality~\cite{Wintersperger_2023}. Results in error mitigation~\cite{Kim2023} and error correction~\cite{acharya2024quantumerrorcorrectionsurface} indicate viable paths towards fault-tolerant quantum computing. 
Also, on the algorithm side, progress is made. For example, Shor's factoring algorithm was recently improved~\cite{regev2024efficientquantumfactoringalgorithm}. In summary, we expect a practical quantum advantage to emerge gradually, not through a single breakthrough, but as part of a broader transition~\cite{scholten2024assessingbenefitsrisksquantum}.

\emph{Benchmarking:}  To effectively use quantum computing today, it is crucial to identify and benchmark problems that align with current hardware structure and limitations~\cite{10.1145/3477206.3477464}. Specific quantum hardware may be suited for particular problem types and structures, making problem-device matching essential. In particular, application benchmarks are instrumental for assessing the end-to-end impact on application-centric metrics.

\emph{Resource Estimation:} Quantum resource estimation bridges the gap between benchmarks and practical implementation roadmaps for quantum algorithms. Using benchmarks alone, it is impossible to accurately assess the quantum resource scale and characteristics needed for specific application scenarios, e.\,g., qubit count, error rates, operation, and measurement speeds. Several tools have emerged, including the Azure Resource Estimator~\cite{beverland2022assessingrequirementsscalepractical},  Qualtran~\cite{misc:qualtran}, and Bench-Q~\cite{zapatabenchq}. Recent research by Krol et\,al.~\cite{Krol2023} utilized the Azure Resource Estimator, revealing significant challenges in achieving near-term quantum advantage for certain algorithms like Grover's search for industrial-scale problems, such as shift scheduling. These findings are consistent with other studies, e.\,g., Hoefler et al.~\cite{10.1145/3571725}.

\emph{Near-term Industry Impact:} In summary, the capabilities of current quantum hardware and algorithms limit its business value. Nevertheless, several quantum-inspired and classical solutions have emerged from projects exploring quantum solutions. For example, Schuetz et al.~\cite{Schuetz_2022} developed a new classical nature-inspired dual annealing algorithm for robot motion planning that is capable of providing high-quality solutions to industrial-scale problems with robot paths of up to 71 seams, resulting in 1 mio. binary variables. The data collection and problem modeling methods developed in such quantum projects can often be successfully transferred to classical solutions delivering business benefits.

Notably, the data acquisition and problem modeling techniques developed in quantum projects can often be effectively transferred to classical solutions, ensuring that significant results can be achieved even with quantum hardware's current computational limitations.

\emph{Future Directions:} Hardware continues to advance, as evidenced by enhanced qubit fidelities~\cite{quantinuum_2024}, error correction breakthroughs~\cite{hong2024entanglinglogicalqubitsbreakeven, Bluvstein_2023}, and new distributed, multi-QPU architectures~\cite{BECK202411}. Emerging quantum-classical architectures, also referred to as HPCQC~\cite{hpcqc} or quantum-centric supercomputing~\cite{alexeev2023quantumcentric},  integrate Quantum Processing Units (QPUs) as accelerators into the high-performance computing stack. While these hardware trends are promising, they require adaptation to algorithms~\cite{conceptual_middleware, saurabh2024quantumminiappsframeworkdeveloping, mantha2024pilotquantumquantumhpcmiddlewareresource}.  These adaptations often involve techniques such as circuit cutting~\cite{10.1145/3579371.3589352}, hardware-software co-design~\cite{10.1145/3477206.3477464}, and error mitigation to optimize performance and reliability.

\subsubsection*{Acknowledgements}
We thank Francesc Sabater Garcia for proof-reading the manuscript. JK and CAR were partly funded by the German Ministry for Education and Research (BMBF) in the project QAI2-Q-KIS under Grant 13N15583. AL was partly funded by the Bavarian State Ministry of Economic Affairs in the project Bench-QC under Grant DIK-0425/03. ME and LM were partly funded by the German Federal Ministry for Economic Affairs and Climate Action (BMWK) in the project QCHALLenge under Grant 01MQ22008D. The authors generated parts of this text with OpenAI's and Anthropic's language-generation models. Upon generation, the authors reviewed, edited, and revised the language.

\bibliographystyle{elsarticle-num}
\bibliography{bibliography, bib_rudi}

\end{document}